\documentclass{IEEEtran}
\usepackage{booktabs}
\usepackage[T1]{fontenc} 
\usepackage{amsmath}
\usepackage{epsfig,endnotes}
\usepackage[colorinlistoftodos,prependcaption,textsize=tiny]{todonotes}
\usepackage{hyperref}

\usepackage{mathtools}
\usepackage{url}
\usepackage[noend]{algpseudocode}
\usepackage{tabularx,colortbl}
\usepackage{multirow}
\usepackage[Symbolsmallscale]{upgreek}
\usepackage{algorithm}
\usepackage{algpseudocode}
\usepackage{mathtools}
\usepackage{url}
\usepackage{authblk}
\usepackage{tablefootnote}
\usepackage{siunitx}
\usepackage{microtype}
\usepackage{float}
\usepackage{graphicx}
\usepackage{adjustbox}
\usepackage{array}

\begin{document}
\title{MobilBye: Attacking ADAS with Camera Spoofing}
\author{
Dudi Nassi, Raz Ben-Netanel, Yuval Elovici, Ben Nassi\\
\textbf{Video Demonstration -} - \url{https://www.youtube.com/watch?v=PP-qTdRugEI&feature=youtu.be}  \\
The Department of Software and Information Systems Engineering\\
Ben-Gurion University of the Negev\\
nassid@post.bgu.ac.il, razx@post.bgu.ac.il, elovici@inter.net.il, nassib@post.bgu.ac.il
}

\maketitle
\thispagestyle{empty}

\section*{Abstract}
\label{section:abstract}
Advanced driver assistance systems (ADASs) were developed to reduce the number of car accidents by issuing driver alert or  controlling the vehicle. In this paper, we tested the robustness of Mobileye, a popular external ADAS. We injected spoofed traffic signs into Mobileye to assess the influence of environmental changes (e.g., changes in color, shape, projection speed, diameter and ambient light) on the outcome of an attack. To conduct this experiment in a realistic scenario, we used a drone to carry a portable projector which projected the spoofed traffic sign on a driving car. Our experiments show that it is possible to fool Mobileye so that it interprets the drone carried spoofed traffic sign as a real traffic sign.

\section{Introduction}

Advanced driver assistance systems (ADASs) \cite{wiki:ADAS} are electronic systems that aid automobile drivers while they are driving. Such systems aim to help drivers by: 1) issuing alerts (e.g., collision avoidance and lane departure alerts) regarding potential threats and 2) recognizing upcoming traffic signs. In order to support the abovementioned  functionalities, These systems use input obtained from multiple sensors including: video camera, LiDAR and radar. ADASs have already become an integral part of the current car generation, and they will provide automated functionalities for the next generation of cars (autonomous vehicles). 

Recently, these systems have attracted increase attention within academia, and the academic community has begun to investigate the systems robustness to various attacks. Recent studies \cite{akhtar2018threat,lu2017standard,petit2015remote,trick-tesla-by-stickers} showed that ADAS alerts and notifications can be spoofed by applying adversarial machine learning techniques to traffic signs, allowing attackers to control the output of the ADAS for their benefit. Application of the methods suggested in these studies exposes drivers that respond to ADAS alerts and notifications and other nearby drivers and pedestrians to variety of risks that can cause accidents. However, the suggested attacks are complicated and require deep understanding of the ADAS used in order to manipulate it. In addition, the nature of the attacks necessitated that the attacker be located near the traffic sign in order to perform the attack. Because of this, the attacks suggested in the recent studies are impractical in real life scenarios. We wonder whether a complicated attack is really needed to manipulate an ADAS?

In this paper, we evaluate the robustness of Mobileye the most popular off-the-shelf ADAS on the market today. It currently provides features like lane departure warnings and traffic sign recognition based on video camera processing. We show that attackers can use a projector in order to inject traffic signs into Mobileye, effectively remote controlling a targeted car according to their wishes. We performed various experiments and assessed the influence of color, shape, projection speed, diameter, and ambient light on the outcome of the attack by mounting a projector onto a drone and injecting traffic signs into the ADAS of a real driving car.

We make the following contributions: First, unlike other studies in this area that trained a classifier and found vulnerabilities to attack the ADAS, we evaluated the practicality of our attack against Mobileye, a real off-the-shelf ADAS. Second, our vector attack doesn't require the attacker to be in the attack location; we present a remote attack that can be executed by a drone. 

\section{Background}
\label{section:background}
Advanced driver assistance systems \cite{wiki:ADAS} were developed to automate, adapt and enhance vehicle systems for safety and improved driving. Most road accidents occur due to human error, and ADASs use input from sensors like video cameras to reduce human error by issuing driver alerts or controlling the vehicle. Such systems have become common in modern cars, with automobile manufacturers integrating these systems in their cars \cite{ADAS-in-mercedes,Driver-assistance-systems-audi}.

There are six levels of automation when it comes to ADASs \cite{wiki:AV}, the highest level (5) is fully automated where the ADAS controls all aspects of the car, and the lowest level (0) where the driver controls all aspects of the car, and can only receive input (in the form of alerts) from the ADAS.

\begin{figure}[h]
\centering
\includegraphics[width=5cm]{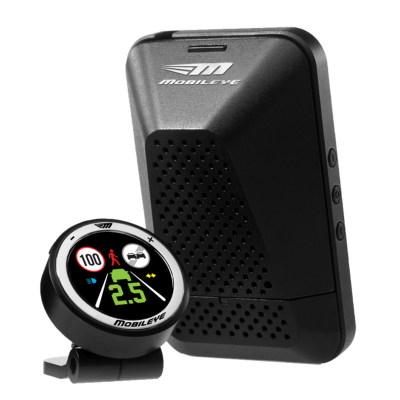}
\caption{Mobileye 630 PRO components}
\label{fig:mobileye}
\end{figure}

Mobileye is an external ADAS that does not provide full automation (Level 5) but rather provides function-specific vehicle automation (Level 0). The Mobileye 630 PRO contains two main components, as can be seen in Figure \ref{fig:mobileye}. The first is a camera, which is installed on the windshield, under the rear view mirror, and the second is a small display which is placed in front of the driver and provides visual and audible alerts about the surroundings as needed. Mobileye has the following features: 

\begin{enumerate}
\item \underline{Lane departure warning}: this is activated when a lane deviation occurs without proper signal notification by the driver; this feature is activated when the driving speed is over 55 kilometers per hour.

\item \underline{Pedestrian collision warning}: this feature notifies the driver of an imminent collision with a pedestrian or cyclist; this feature can only be used during daylight and is activated when the driving speed is under 50 km/h (this feature can be configured to be up to 70 km/h). 

\item \underline{Forward collision warning}: This feature notifies the driver about rear-end collisions with any type of vehicle. 

\item \underline{Headway monitoring and warning}: This feature notifies the driver when there is an unsafe distance between the driver's vehicle and the vehicle ahead of it, this feature is activated when the driving speed is over 30 km/h. 

\item \underline{Intelligent high beam control}: This turns the high beams on and off depending on the light and relative distance from other vehicles, a feature which is only used at night. 

\item \underline{Recognizing upcoming traffic signs}: Recognizing and reading traffic signs (speed limit, entering a highway, etc.).

\end{enumerate}

\section{Related Work}
\label{section:related-work}
In this section, we describe related work on attacks against ADASs and provide an overview of adversarial attacks. The computer vision classifier is an integral ADAS component which is used to detect traffic signs from a video stream in an ADAS. Many of these classifiers are trained using deep learning techniques. Several studies created adversarial instances to trick such deep learning classifiers and showed that this type of classifier is vulnerable to spoofing attacks. \textit{Petrakieva et al.} \cite{akhtar2018threat} demonstrated how perturbations that are often too small to be perceptible to humans can fool deep learning models. \textit{Sitawarin et al.} \cite{sitawarin2018darts} showed that they could embed two traffic signs in one traffic sign with a dedicated array of lens that causes a different traffic sign to appear depending on the angle of view. \textit{Eykholt et al.} \cite{eykholt2017robust} and \textit{Lu et al.} \cite{lu2017standard} showed that physical artifacts (e.g., stickers, graffiti) misled computer vision classifiers. In the abovementioned studies, the researchers only trained dedicated models by themselves and identified instances that could exploit them using white-box techniques. Furthermore, the researchers did not show the effectiveness of the attack against an off-the-shelf ADAS. In contrast, we demonstrate our attack against the Mobileye system and mislead it so it recognizes spoofed traffic signs using black-box techniques.

Attacks against ADAS are not, however, limited to misleading the classifier using an adversarial traffic sign. \textit{Petit et al.} \cite{petit2015remote} presented two attack vectors against car's sensors such as LiDAR and cameras. They were able to show that: 1) a laser directed at the camera can destroy the optical sensor permanently, and 2) LiDAR's output can be spoofed using infrared light. \textit{Yan et al.} \cite{yan2016can} demonstrated various spoofing attacks against a camera, ultrasonic sensor, and radar that can cause Tesla's Model S to misperceive the distance of nearby obstacles. However, it is not possible to perform the suggested attacks \cite{petit2015remote, yan2016can} on a driving car due to the complexity of the attacks because: 1) they require deploying devices at specific ranges from the attacked car, and 2) the attacker must connect the hardware directly to a driving autonomous car which can be a major challenge due to the driving speed. A recent study \cite{trick-tesla-by-stickers} misdirected an autopiloted vehicle, taking it in the wrong direction. The authors placed interference patches (small stickers) on the ground at two way route, causing the vehicle to turn in to the opposite lane. In this case, the attacker must physically put the stickers on the road; in contrast, our attack vector doesn't require the attacker to be on site, since the drone can be deployed remotely.

Other famous attacks against cars that are not related to our work are \cite{greenberg2017hackers, checkoway2011comprehensive}, which were based on compromising the firmware of the car or an internal device. Our attack model is much lighter than these attacks, since it does not require us to hack to those systems.

\section{Threat Model}
\label{section:threat-model}
We consider an attacker as any malicious entity with the aim of attacking a driving car. The attacker can inject spoofed traffic signs into Mobileye using a portable projector mounted on a drone. The attacker's goals can be to: 1) harm or manipulate the car of a specific victim, or 2) cause environmental chaos (e.g., harm multiple cars in a specific region such as a city, neighborhood, highway, etc.).

In this paper, we present a means of executing this threat model, using a drone equipped with a portable projector as our injection method. Mobileye (Figure \ref{fig:mobileye}) is the sensor we are going to mislead by projecting a traffic sign that does not fit the surroundings and risks the driver/car that follows the injected traffic sign. 

In our study we show focus on the traffic signs recognition.
\section{Mobileye Analysis}
\label{section:analysis}
In order to execute our suggested threat model (described in Section \ref{section:threat-model}), we focus on Mobileye's traffic sign recognition features. In the following subsections, we learn the effect of environmental factors (ambient light, distance) on the result of the attack. In addition, we test the robustness of the Mobileye for classifying traffic signs that do not exist. 

\subsection{Experimental Setup}
\label{subsection:exp-set}
\begin{figure}[h]
\centering
\includegraphics[width=8cm]{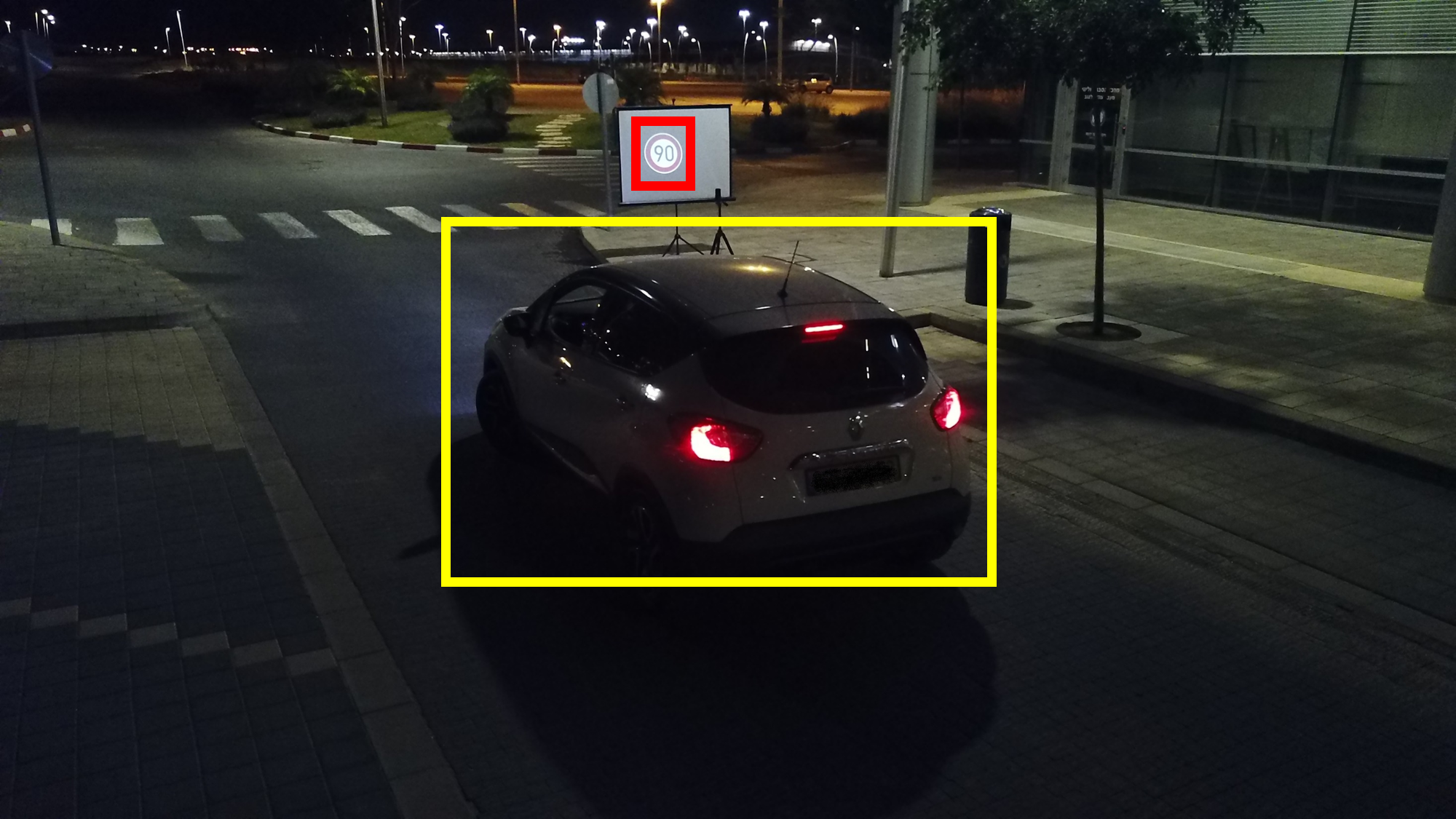}
\caption{Experimental setup: the projected sign is boxed in red, and the attacked vehicle is boxed in yellow}
\label{fig:exp_set}
\end{figure}
In this subsection, we describe the setup for the experiments performed. For convenience we used a white projector screen, in order as screen for the projected traffic sign. A portable projector was used to provide the sign's content. The injection method, as described in Section 4, is comprised of the projector and screen. The portable projector was placed on a tripod about 2.5 meters from the screen and projected a traffic sign onto the center of the screen; while the sign was projected in this way we drove the car (a Renault Captur equipped with Mobileye 630 PRO) in a straight line at a speed of approximately 25 km/h. Figure \ref{fig:exp_set} presents an illustration of our experimental setup. 

\subsection{Influence of the Projected Sign's Diameter}
\begin{figure}[h]
\centering
\includegraphics[width=8cm]{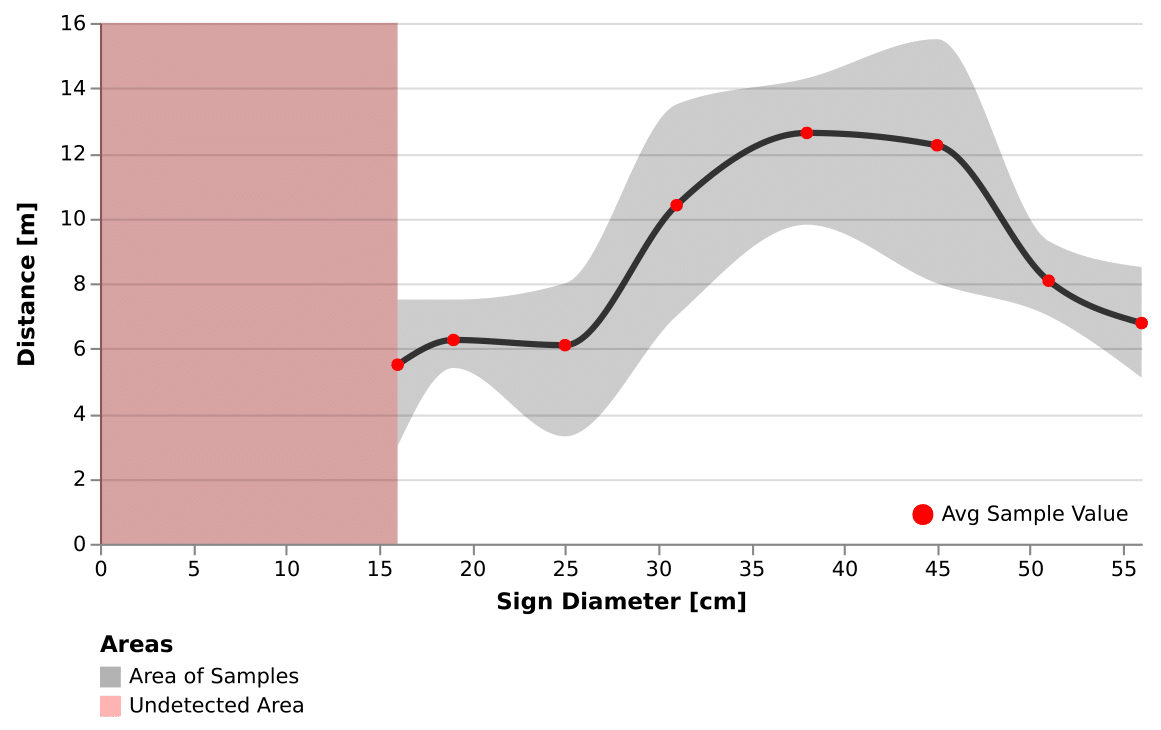}
\caption{Influence of the sign's diameter}
\label{fig:sign_diam}
\end{figure}

\subsubsection{Experimental Setup}
In this case, we investigate whether the size of the projected sign influences the distance from which the Mobileye 630 PRO's sensor can detect the projected sign. We repeated the experiment five times, projecting a different sized sign each time, and calculated the average detection distance.
\subsubsection{Results}
Figure \ref{fig:sign_diam} presents the results of this experiment. As can be seen, if the sign is too small (less than 16 cm in diameter) the Mobileye 630 Pro sensor didn't detect it at all. The red dots in the graph symbolize the average distance at which we managed to mislead the sensor, and the grey area shows the range of the entire samples set.
\subsubsection{Conclusion}
The diameter range is wide and provides a lot of room for error when projecting a traffic sign. Based on our measurements, the distance can range from approximately 5-16 meters.

\subsection{Influence of the Color of the Projected Sign}

\begin{figure}[h]
\centering
\includegraphics[width=8cm]{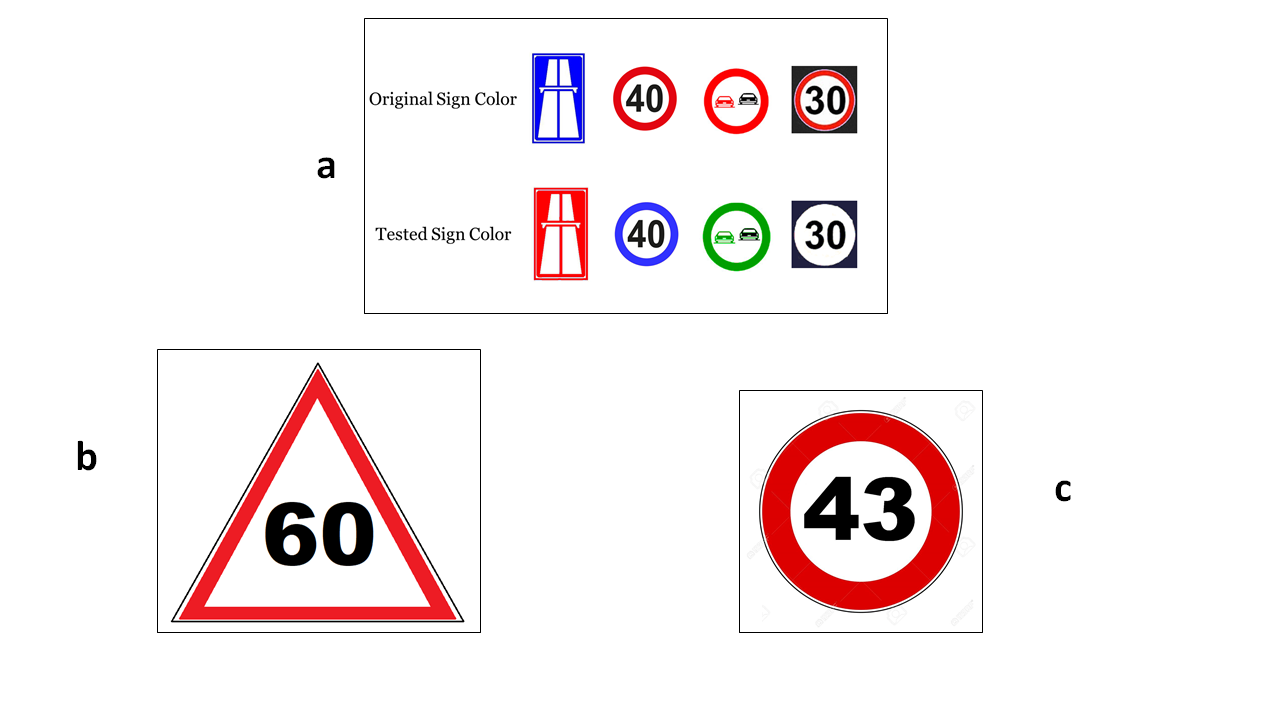}
\caption{(a) examples of different colored traffic signs, (b) an example of a different traffic sign shape, (c) an example of an incorrect or unrecognized speed limit value}
\label{fig:exp}
\end{figure}

\subsubsection{Experimental Setup}
Here we assess whether the Mobileye 630 PRO sensor is sensitive to the color of the sign. We tried various colors as seen in (Figure \ref{fig:exp}a). First we projected the sign with its true colors, and then we verified that the Mobileye 630 PRO sensor managed to recognize the sign. Next, we projected the same sign but this time with a color scheme which is different from the real one.
\subsubsection{Results}
We could see, quite quickly, that Mobileye is not sensitive to color, since all of the signs tested managed to mislead the sensor (even the black and white speed limit sign seen in, Figure \ref{fig:exp}a). 
\subsubsection{Conclusion}
Based on these results we conclude that the Mobileye 630 PRO sensor only considers the shape of a sign when trying to classify the sign's content.

\subsection{Influence of the Projected Sign's Shape}
\subsubsection{Experimental Setup}
In this case, we evaluate whether the Mobileye 630 PRO sensor can detect signs which do not take the form of their original shape. For this experiment we simply took a speed limit sign and modified its shape (as seen in Figure 4b). This experiment was binary, i.e., we only wanted to know if the sign can be detected or not.
\subsubsection{Results}
We utilized a total of seven different shapes (a triangle, rectangle, pentagon, and hexagon, as well as three other more unusual shapes - a star, arrow, and random polygon). The Mobileye 630 PRO's sensor was unable to detect any of these shapes. 
\subsubsection{Conclusion}
We can conclude the Mobileye system considers just the shape of the sign and isn't fooled by unknown shapes.

\subsection{Influence of Ambient Light }
\label{subsection:Ambient}
\begin{figure}[h]
\centering
\includegraphics[width=8cm]{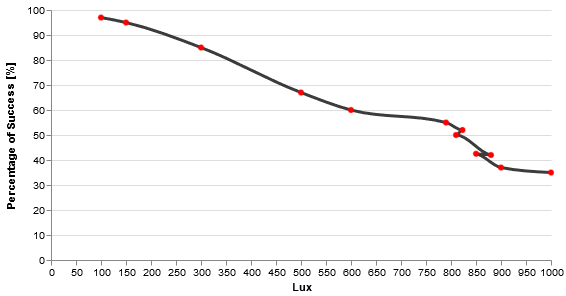}
\caption{Influence of Ambient Light}
\label{fig:ambient_light}
\end{figure}
 
\subsubsection{Experimental Setup}
In this case, we tested the effect of ambient light, utilizing 20 samples (drives) from every hour of the day to check our injection success rate.
\subsubsection{Results}
Figure \ref{fig:ambient_light} presents the results of this experiment; success is considered a sample (drive) in which Mobileye recognized the projected sign.
\subsubsection{Conclusion}
Based on our analysis of these results we can conclude that is possible to inject a spoofed traffic sign at all hours of the day, but performance is best later in the day (in the evening and at night). One thing that should be considered with regard to ambient light is the equipment used, since the opacity of the projected sign depends on the ambient light as well as the projector used (a better success rate may be achieved with a better projector). 

\subsection{Influence of the Speed of the Projection Time}
\subsubsection{Experimental Setup}
Here we assessed the speed of the projection time that is needed to fool the system. We conducted a few experiments that measured the amount of time required for injection.
\subsubsection{Results}
We discovered that a projection speed of 100 ms is sufficient for fooling the system. We were unable to fool the system with faster projection speeds probably due to the frame per second rate of the optical sensor of the Mobileye.
\subsubsection{Conclusion}
The fast projection speed causes the attack vector on the target to be very easy to inject and doesn't require staying for to long.

\subsection{Influence of the Number on the Projected Sign}
\label{subsection:numbers}
\subsubsection{Experimental Setup}
In this case, we investigate whether the Mobileye 630 PRO sensor can also detect speed limit signs with speed values that are not used in the real world (e.g. Figure \ref{fig:exp}.c).
\subsubsection{Results}
Table 1 presents the results of this experiment.

\begin{table}
\caption{Detection of incorrect speed limit signs,(Left: speed limit on the projected sign. Right: the detected speed limit, as shown on  the Mobileye display, X means no detection) }~\label{tab:results}
 \begin{tabular}{||c c||} 
 \hline
Projected Speed limit & Detected Speed limit \\ [0.5ex] 
 \hline\hline
0 & X  \\ 
 \hline
 1 & X  \\
 \hline
  2 & X  \\ 
 \hline
  3 & X  \\ 
 \hline
  4 & X  \\ 
 \hline
  5 & 5  \\ 
 \hline
  6 & X  \\ 
 \hline
  7 & X  \\ 
 \hline
  8 & X  \\ 
 \hline
  9 & X  \\ 
 \hline
  27 & X  \\ 
 \hline
  43 & X  \\ 
 \hline
  69 & 60  \\ 
 \hline
 71 & 70  \\
 \hline
 88 & 80 \\
 \hline
  150 & X  \\ 
 \hline
  160 & X  \\ 
 \hline
  170 & X  \\ 
 \hline
  180 & 110  \\ 
 \hline
   190 & 110  \\ 
 \hline
 200 & X  \\ [1ex] 
 \hline
\end{tabular} 
\end{table}
\subsubsection{Conclusion}
Based on these results, we can conclude that incorrect speed limit signs are effective at misleading the system. The system do not ignore them and classify them as other similar traffic signs.

\section{Attacking a Car While Driving}
\label{section:attacking a Car While Driving}
\begin{figure*}[tbp]
\centering   
\includegraphics[width=0.34\textwidth]{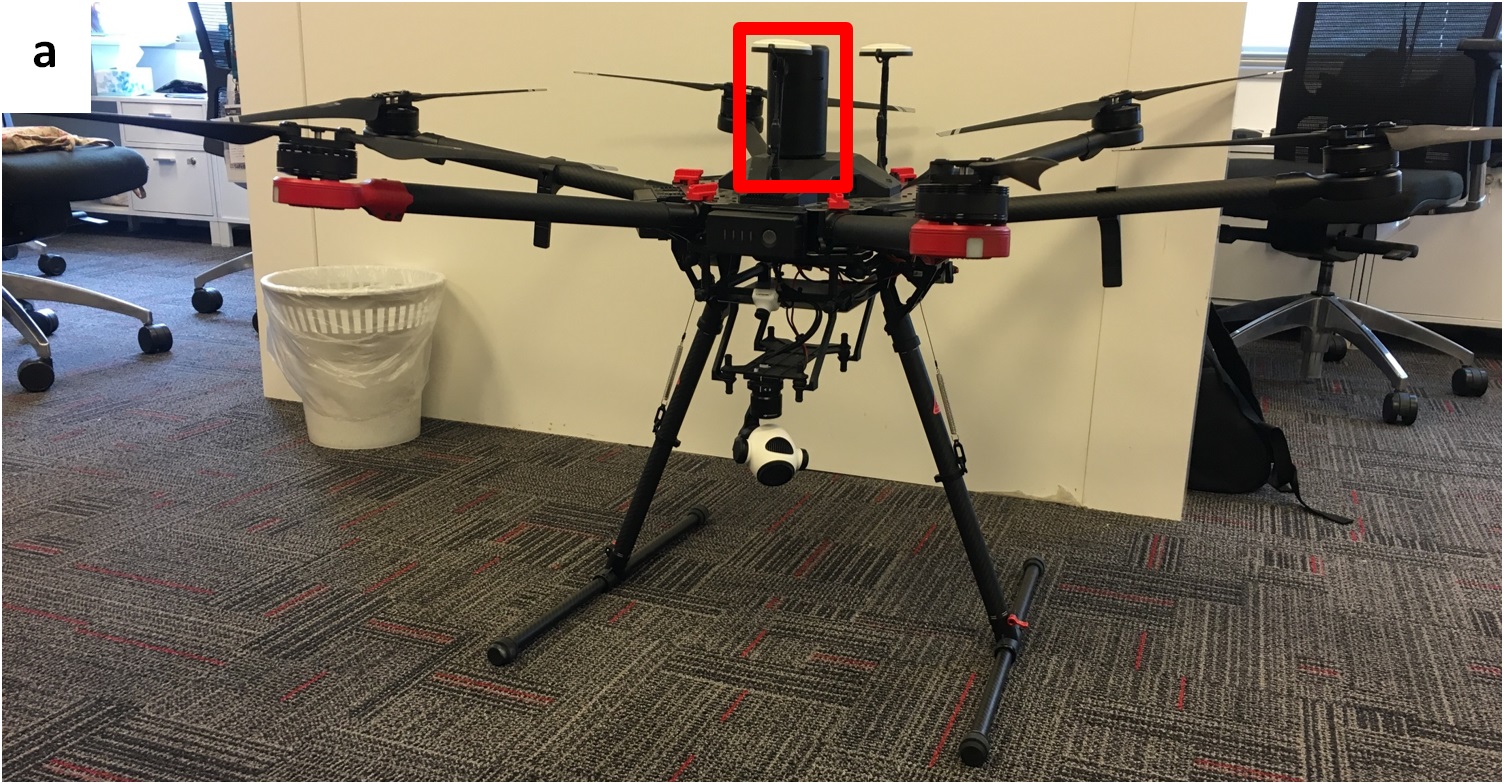}
\includegraphics[width=0.34\textwidth]{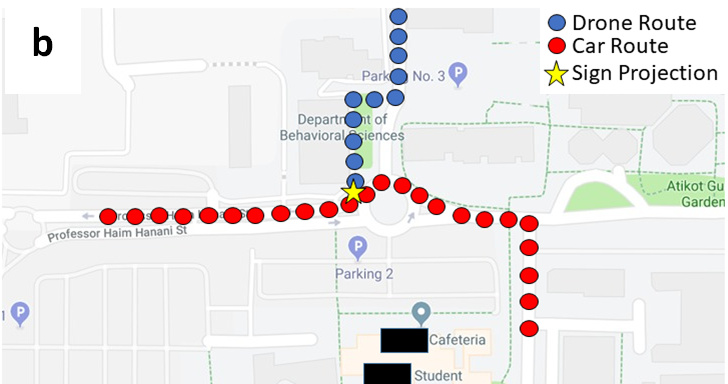}
\includegraphics[width=0.3\textwidth]{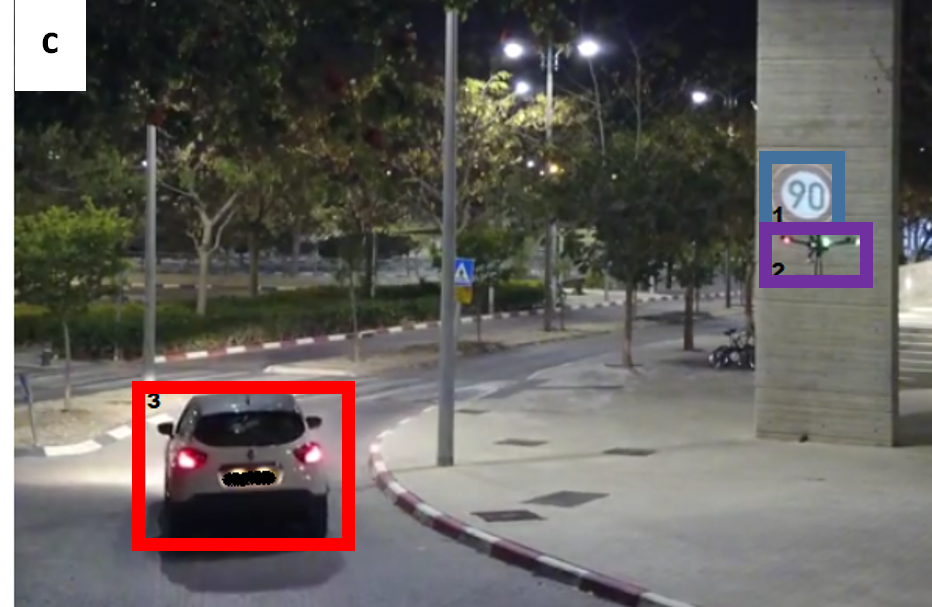}
\caption{Attacking a car while driving (a) the drone with the projector used in our experiments, (b) visualization of the threat model implementation, (c) the moment of the attack (the projected sign is boxed in blue, the attacker's drone is boxed in purple, and the victim's car is boxed in red).}
\label{fig:Attack-demonstration}
\end{figure*}
\begin{figure}[h]
\centering
\includegraphics[width=8cm]{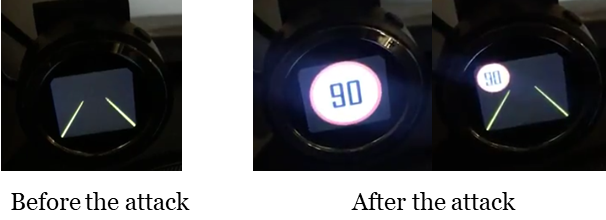}
\caption{Mobileye display before and during the attack. }
\label{fig:after_before}
\end{figure}
In this section, we demonstrate how attackers can apply the attack remotely using drone.
\subsubsection{Experimental setup}
We mounted a portable projector on a drone (DJI Matrice 600) (Figure \ref{fig:Attack-demonstration}a). In this experimental setup our car (a Renault Captur equipped with Mobileye 630 PRO) was driven in an urban environment as the attacker operated a drone, positioning the drone so the spoofed speed limit sign can be injected into the Mobileye sensor. The routes can be seen in Figure: \ref{fig:Attack-demonstration}b. At the starred location on the map, the attacker projected the incorrect speed limit sign (seen in Figure \ref{fig:Attack-demonstration}c), managing to mislead the Mobileye sensor which recognized the sign as a 90 km/h speed limit sign (see Figure \ref{fig:after_before}). The implemented attack vector can be seen in an uploaded video of the attack \footnote{\label{car} \url{https://www.youtube.com/watch?v=PP-qTdRugEI&feature=youtu.be}}.
\subsubsection{Results}
We managed to fool Mobileye so it classified the speed limit as 90 km/h when the speed limit for a city road is only 30 km/h.

\section{Countermeasures}
\label{section:countermeasures}
In this section, we discuss countermeasure methods that can be used to prevent computer vision classifiers used by ADASs from being misled by an attacker. The suggested measures are passive which means that they 1) don't require any source of power for transmitters, and (2) prevent all kinds of visual spoofing attacks, including projected traffic signs and image perturbations that mislead the system to consider (classify) a sign as a real sign.
\begin{figure}[h]
\centering
\includegraphics[width=8cm]{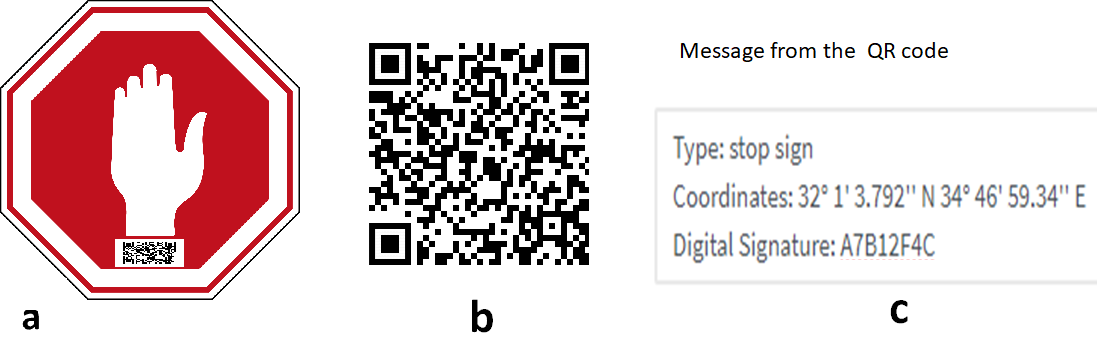}
\caption{Traffic sign authentication countermeasure. (a) example of the traffic sign with the QR code, (b) example of the QR code included in the message sent to the car, (c) the message extracted from the QR code (including the signature). }
\label{fig:QR-example}
\end{figure}

We found that inserting a QR code into traffic signs is the most effective way of preventing the Mobileye sensor from being misled by a spoofed traffic sign; more specifically, in this countermeasure method, the QR code contains a signed message to the approaching car, informing the car of the traffic sign ahead, and its type, coordinates, and digital signature for authentication during the recognition process. In this way: 1) the accurate location of the sign and the sign type can be authenticated, and 2) the digital signature is used to verify that the traffic sign is not fake. This solution can, however, be vulnerable to GPS spoofing and can cause the Mobileye system to recognize the current sign as another sign located elsewhere due to the change of coordinates (an example of a sign and message can be seen in Figure \ref{fig:QR-example}).

\begin{table}
\caption{Advantages and disadvantages for each countermeasure method}~\label{tab:adv/dis}
\begin{tabular}{ | m{2.5cm} | m{2.5cm}| m{2.6cm} | }
\hline
Method & Advantages & Disadvantages \\ \hline
Image authentication & Allow verification of traffic signs & Vulnerable to GPS spoofing \\ \hline
3D signs & Mitigate projection attack & Vulnerable to replay attacks that require physical approach \\ \hline
Social navigation application(Waze) & Verify against updatable database & Require Internet connectivity and vulnerable to GPS spoofing \\ \hline
\end{tabular}
\end{table}

3D traffic signs can also be used to eliminate the threat of projected traffic signs. Since our attack model (2D projection) is aimed at the video camera which can't distinguish between 2D and 3D signs (see Subsection \ref{subsection:exp-set}), the obvious way to counter the attack is to use sensors that can detect 3D objects (such as LiDAR); however, an attacker could make 3D sign that contains the content desired by the attacker and physically place it everywhere.

Our last countermeasure suggestion involves the use of a social navigation application (such as Waze) that can map all of the traffic signs and create a traffic sign database; in this way the ADAS can verify the traffic signs using the database. This solution requires Internet connectivity for updating the traffic signs using information sent from other drivers, making it vulnerable to cyber-attacks and GPS spoofing.

Our suggestions can be implemented in a short period of time, whereas the next generation of traffic signs will take much longer. Next generation "car-to-car communication" will include transmitters in traffic lights and signs, which will transmit signals to the car, effectively providing the car with instructions; the car will then operate in accordance with the instructions. Adaptation of the overall transportation environment in light of these changes will take some time; until then the countermeasures presented may be helpful.

\section{Responsible Disclosure}
\label{section:Diclosure}

We are currently in the middle of the process of responsible disclosure. We sent our findings to Mobileye and Tesla and we are waiting for their response.

\section{Discussion}
\label{section:discussion}

In this section, we discuss our research which provides new insights into the risks associated with both autonomous and nonautonomous vehicles. In this paper, we showed how easy is to maliciously fool ADASs, such as Mobileye. These systems provide the data need for autonomous driving, and our sign spoofing attack presents a major threat to autonomous vehicles which rely on these systems. In addition, our discussion of related work revealed many vulnerabilities, causing us to wonder whether AV development should continue without first addressing the security issues mentioned. We know that in the future new communication methods (vehicle-to-vehicle/V2V and vehicle-to-infrastructure/V2I communications) to improve driving and safety will be implemented. V2V communication will be used for exchanging data between vehicles, including information about the application of brakes, speed, location, live threats from the road, etc., while V2I communication will be used for exchanging data between vehicles and the environment (e.g., recognition of traffic lights and traffic signs); in both cases, information exchange will take place via radio frequencies (Wi-Fi). These next generation systems will not be vulnerable to computer vision attacks like traffic sign spoofing and ADAS manipulation, because projection or perturbations of the signs won't be relevant - data will flow over radio frequencies - however the systems of the future will be vulnerable to new attack vectors (e.g., GPS spoofing, hacking Wi-Fi transmitters and mimicking transmitters, and jamming). It is our hope that our research will contribute to improved driving and safety now and in the future.

\bibliographystyle{IEEEtran}
\bibliography{IEEEabrv,main}

\footnotesize 
\Urlmuskip=0mu plus 1mu\relax

\end{document}